# Materials processing with tightly focused femtosecond vortex laser beams

Cyril Hnatovsky[1], Vladlen G. Shvedov[1,2], Wieslaw Krolikowski[1], Andrei V. Rode[1]

*[1]Laser Physics Centre and [2]Nonlinear Physics Centre,*

*Research School of Physics and Engineering, The Australian National University, Canberra ACT 0200, Australia*

*Corresponding author: chn111@physics.anu.edu.au

**Abstract**: This letter is the first demonstration of material modification using tightly focused femtosecond laser vortex beams. Double-charge femtosecond vortices were synthesized with the polarization-singularity beam converter described in Ref [1] and then focused using moderate and high numerical aperture optics (viz., NA = 0.45 and 0.9) to ablate fused silica and soda-lime glasses. By controlling the pulse energy we consistently machine high-quality micron-size ring-shaped structures with less than 100 nm uniform groove thickness.

Currently, ultra-short pulse lasers are extensively used both for precision materials processing and in fundamental studies dealing with the interaction of high intensity radiation with matter. By shaping the laser beam (e.g., its spatial intensity and temporal profile) it is possible to control the light-matter interaction. For instance, high intensity optical vortex pulses provide an opportunity to investigate the effects of the optical angular momentum on atomic or molecular systems [2]. Recently it has been suggested that ablation of Ta (i.e., tantalum) targets with nanosecond vortex pulses provides clearer and smoother processed surfaces than the ablation performed with non-vortex pulses [3]. Femtosecond pulses however offer much higher precision in micromachining due to the absence or very limited heat wave and shock wave affected volume [4, 5]. Unfortunately, structuring and modification of materials using femtosecond vortex pulses poses a significant challenge due to technical difficulties hindering the synthesis of high-power broadband laser vortex beams. The traditional methods of generating femtosecond optical vortices with spiral phase plates and (computer generated) holograms are inherently chromatic and therefore require the introduction of correcting elements in order to compensate topological charge dispersion occurring in polychromatic pulses. Uncompensated charge dispersion leads to a poor quality of the beam and the impossibility of using it for precision laser materials processing [6]. The compensation schemes, however, come at the expense of making the beam-shaping setups significantly more complicated and sensitive to alignment and remain either bandwidth-limited, or low throughput, or unsuitable for the operation with high-intensity laser pulses due to high energy dissipation inside the optical components ([7] and references therein). So far, none of the proposed compensation techniques have been tested or used in laser micromachining applications.

Recently we have demonstrated that high-quality femtosecond vortex beams can be easily synthesized by using polarization singularities associated with the beam propagation in birefringent crystals [1]. In this Letter we employ this powerful technique to generate double-charge femtosecond laser vortex beams and use them for reproducible sub-micrometer structuring of fused silica ($SiO_2$) and soda-lime glass samples. To the best of our knowledge, this is the first report on materials processing using tightly focused femtosecond vortex beams.

Polarization singularities can be created when converging/diverging light propagates through an anisotropic medium [8, 9]. In Ref [1] it was shown that when a circularly polarized femtosecond Gaussian beam $\mathbf{E}^{\pm in} = (E/\xi)\exp(-r^2/(w_0^2\xi))\mathbf{c}^{\pm}$ with a waist radius $w_0$ propagates along the optical axis of a uniaxial crystal, its field (i.e., $\mathbf{E}^{\pm in}$) is converted into a superposition of two polarization states with opposite handednesses:

$$\mathbf{E}^{\pm in} \Rightarrow \mathbf{E} = 0.5(\mathbf{c}^{\pm}(G^o + G^e) - \mathbf{c}^{\mp}((r^2 + w_0^2\xi_o)G^o - (r^2 + w_0^2\xi_e)G^e)\exp(\pm 2i\varphi)r^{-2}) \quad (1)$$

In the above expressions: $G_{o;e} = (E/\xi_{o;e})\exp(-r^2/(w_0^2\xi_{o;e}))$; $\xi = 1 + iz\lambda/(\pi w_0^2)$; $\xi_o = 1 + iz\lambda/(\pi n_o w_0^2)$; $\xi_e = 1 + iz\lambda n_o/(\pi n_e^2 w_0^2)$; $\mathbf{c}^{\pm} = (\mathbf{e}_r \pm i\mathbf{e}_\varphi)\exp(\pm i\varphi)/\sqrt{2}$; $n_o$ and $n_e$ stand for the ordinary and extraordinary refractive index, respectively. The state with the handedness opposite to that of the input beam carries a double-charge optical vortex ($l = \pm 2$), whereas the state with the handedness coinciding with that of the input beam represents a non-vortex beam (Fig. 1(a)). After the crystal, the wave propagating in free space becomes a superposition of independent solutions of the scalar wave equation $(\nabla_\perp^2 + 2ik_0\partial_z)\Psi = 0$ and the vortex solution for $l = +2$ can be written as:

$$\Psi^{l=2} = \Psi_{z+\delta} - \Psi_{z-\delta} \quad (2)$$

where $\Psi_{z\pm\delta} = (r^2 + w_0^2\xi_{z\pm\delta})G_{z\pm\delta}\exp(2i\varphi)r^{-2}$; $G_{z\pm\delta} = (E/\xi_{z\pm\delta})\exp(-r^2/(w_0^2\xi_{z\pm\delta}))$; $\xi_{z\pm\delta} = 1 + i(z\pm\delta)\lambda/(\pi w_0^2)$; and $\delta$ denotes the axial shift of the waists of the ordinary and extraordinary beams with respect to each other due to double refraction in the crystal. For paraxial beams the shift is approximated by $\delta = d(n_o^2 - n_e^2)/(2n_e^2 n_o)$, where $d$ is the thickness of the crystal along the beam propagation direction $z$. Focusing of such beams was studied in detail in Ref [10].

In our experiments we used the output beam of a Clark-MXR femtosecond Ti:Sapphire amplifier with a central wavelength at $\lambda = 780$ nm. In Figure 1(a) showing the experimental setup for the generation of double-charge femtosecond vortex beams i) the first achromatic quarter-wave plate converts the linearly polarized Clark-MXR's beam into a circularly polarized beam, ii) this beam is defocused with a negative lens L1 (-50 mm) and after propagation along the optical axis of a 10 mm-long $c$-cut $Ca_2CO_3$ crystal is collimated with a positive lens L2 (+125 mm). The second quarter-wave plate together with a polarization beamsplitter PBS separates the emerging double-charge vortex beam from the non-vortex beam [1]. Therefore, polarization of the generated vortex beam is orthogonal to that of the incident laser beam. The presented optical scheme also allows one to synthesize femtosecond vortices with opposite topological charges by simply reversing the handedness of the input polarization with the first quarter-wave plate. For consistency, the topological charge in our experiments was $l = +2$. In the cross-section the generated vortex beam generally consists of a series of concentric rings, as shown in the inset of Fig. 1(a). The number of rings is determined by the input beam diameter at L1, the optical power of L1 and the crystal's parameters (see (1)). In our experiments the beam expander comprised of L1 and L2 was adjusted to allow only the first bright ring of the vortex beam to enter the objective O1 (i.e., the first dark ring coincides with the entrance aperture of O1) to be used for material modification. The simulations in Fig. 1(b) also represent this situation.

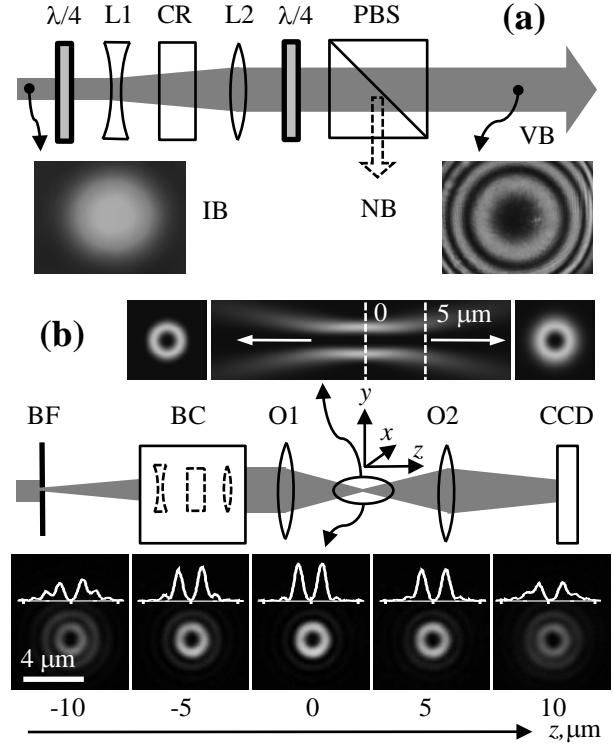

Fig. 1. (a) setup for the generation of double-charge femtosecond vortex pulses. λ/4 - achromatic quarter-wave plate; L1 - negative lens; CR - uniaxial crystal; L2 - positive lens; PBS - polarization beamsplitter; IB and VB - CCD images (6.4 x 4.2 mm$^2$) of incident and vortex beam, respectively; NB - non-vortex beam which is removed from the system. (b): simulated (top) and experimental (bottom) intensity distributions of a focused polarization-singularity vortex beam in the *xz*- and *xy*-plane. O1 and O2 - focusing and imaging objectives; BC - beam converter shown in (a); BF - beam filter (viz., an aperture ~1 mm in diameter located ~5 m from BC). CCD images show the intensity distribution in the *xy*-plane at the focus of O1 with NA = 0.45.

To compare the simulated and actual behavior of a femtosecond vortex beam, we mapped the intensity distribution in the focal region of O1 with an imaging objective O2 having a numerical aperture (NA) significantly higher than that of O1, viz., NA = 0.9 (Nikon M Plan 100x) vs. NA = 0.45 (Olympus LUCPlan FLN 20x) (see Fig. 1(b)). The intensity distributions in the *xy*-plane presented in the bottom panel of Fig. 1(b) were obtained by moving O1 in 5 µm steps along the laser beam propagation direction *z* while keeping O2 in a fixed position. The images were captured with a CCD camera and further analyzed using ImageJ 1.34u software. The analysis shows that the peak-to-valley intensity variation along the maximum of the first bright ring, which according to our measurements has a diameter of ~2 µm at the focus, does not exceed 15%. The average intensity along the first ring at the focus (i.e., $z = 0$) is ~1.2 and ~2 times higher than those measured at $z = \pm 5$ µm and $z = \pm 10$ µm, respectively. Based on these numbers, the 'confocal parameter' of the generated beam is ~20 µm, i.e., it is ~3 times larger than the confocal parameter of a Gaussian beam with a 2 µm waist diameter. The observed elongated tubular focus agrees with the simulations shown in Fig. 1(b), which were performed for O1 with NA = 0.45 and $\lambda = 780$ nm.

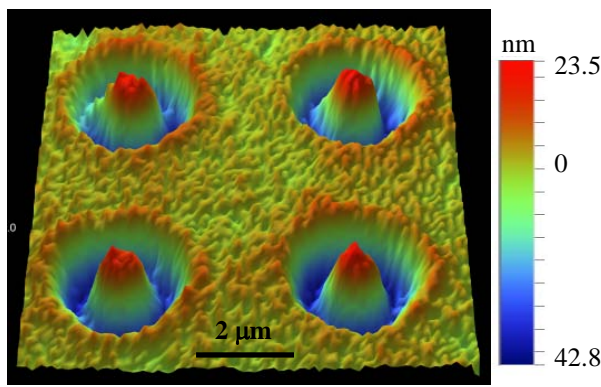

Fig. 2. Ablation of $SiO_2$ with single double-charge 120 nJ femtosecond vortex pulses using NA = 0.45 focusing optics. Ablation craters are separated by 5 µm.

Two focusing regimes have been explored in our experiments on material modification using 200 fs (full width at half maximum after BC) linearly polarized double-charge vortex pulses. In the first case, O1 had a moderate NA of 0.45 (Olympus LUCPlan FLN 20x). The laser beam was focused onto the surface of a fused silica $SiO_2$ sample. Figure 2 shows topographic images of the ablation craters produced with a single pulse at 120 nJ (after O1) which were recorded with a Wyko NT9100 surface profiler providing a 0.45 μm lateral resolution and a sub-nanometer vertical resolution. The depth of the ablation craters is ~40 nm and the profile is uniform along the grooves. The lateral dimensions of the craters agree with the intensity distribution in the focal region of O1 shown in Fig. 1(b).

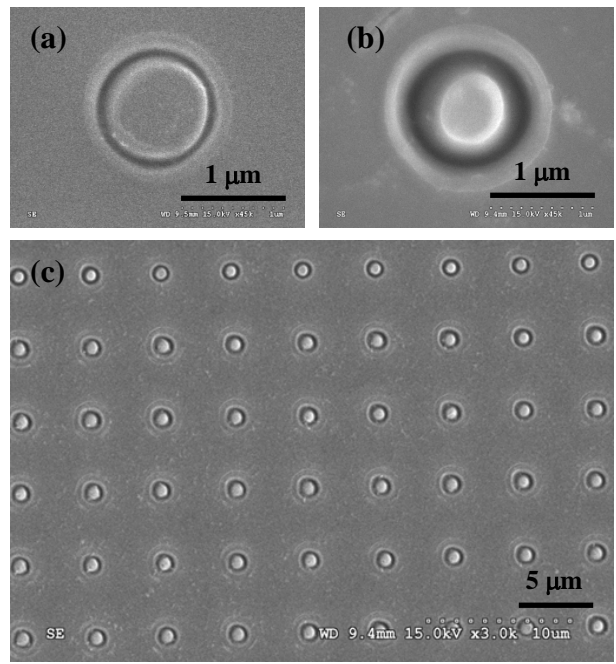

Fig. 3. Ablation of soda-lime glass with single double-charge femtosecond vortex pulses using NA = 0.9 focusing optics. Ablation craters are separated by 5 μm. (a) and (b) are ablation craters produced at a pulse energy of 80 nJ and 150 nJ, respectively. (c) shows a region of a 200 x 200 μm$^2$ array produced at a pulse energy of 150 nJ.

In the second set of experiments samples of soda-lime glass were irradiated using O1 with a high NA of 0.9 (Nikon M Plan 100x). After the irradiation the samples were coated with ~7 nm of Pt and examined under a Hitachi 4700S scanning electron microscope (SEM). The SEM images presented in Fig. 3(a) and 3(b) show the ablation craters produced with single 80 nJ and 150 nJ pulses, respectively. From Fig. 3(a) one can see that i) the diameter of the ablation crater is ~1 μm, i.e., it is one half of the value which was obtained in focusing with NA = 0.45 and ii) the width of the ablated annular groove is less than 100 nm. In Fig. 3(b) the ablation signature is much more pronounced, but it still preserves an annular shape and has sharp and clean edges. It is also noteworthy that ablation of material performed under these conditions is highly reproducible (Fig. 3(c)). The last feature is due to the extended depth of focus of the generated beams, which makes the ablation process much less sensitive to the tilt of the sample and the cosine errors of the translation stages.

In summary, we have demonstrated precision materials processing with femtosecond vortex beams. These beams were generated using the beam converter based on light propagation inside uniaxial birefringent crystals and focused to diffraction-limited spots using both moderate and high NA optics to modify materials on a sub-wavelength scale in a highly controllable fashion. The extended depth of focus of polarization-singularity vortex beams makes the laser ablation process less susceptible to the sample misalignment and translation stage errors and can also significantly increase the ratio of cutting depth to cutting width compared to that achieved with Gaussian beams.

We acknowledge the financial support from the National Health and Medical Research Council of Australia and the Australian Research Council. We also acknowledge Dr. K. Vu for his assistance in recording depth profiles of the ablated samples.

**References (with titles)**